\documentclass[twocolumn,showpacs,showkeys]{revtex4-1}
\usepackage{graphicx}

\begin{document}
\title{Zenith angle distribution of cosmic ray showers measured with the Yakutsk array\\ and its application to analysis of arrival directions in equatorial coordinates}
\author{A.A. Ivanov}
\email{ivanov@ikfia.ysn.ru}
\affiliation{Shafer Institute for Cosmophysical Research and Aeronomy, Yakutsk, 677980, Russia}
\date{\today}
\pacs{96.50.S-, 96.50.sd}
\keywords{cosmic rays, extensive air showers}

\begin{abstract}
The Yakutsk array dataset in the energy interval $(10^{17},10^{19})$ eV is revisited in order to interpret the zenith angle distribution of an extensive air shower event rate of ultra-high energy cosmic rays. The close relation of the distribution to the attenuation of the main measurable parameter of showers, $S_{600}$, is examined. Measured and expected distributions are used to analyze the arrival directions of cosmic rays on an equatorial map including the energy range below $10^{18}$ eV, which was previously avoided due to the reduced trigger efficiency of the array in the range. While the null hypothesis cannot be rejected with data from the Yakutsk array, an upper limit on the fraction of cosmic rays from a separable source in the uniform background is derived as a function of declination and energy.

\end{abstract}

\maketitle

\section{Introduction}
The zenith angle distribution of extensive air showers (EASs) of cosmic rays (CRs) has been a target of investigations since the very beginning of EAS measurements. The results of the classical period of observations were summarized in a monograph \cite{Hayakawa} where the interconnection between the omnidirectional frequency of showers and attenuation of EASs in the atmosphere was analyzed using the \textit{Gross transformation}. At that time, EAS density had to be measured without knowing the incident direction.

Modern EAS arrays measure both the sizes and incident directions of showers. Examples of zenith angle distributions of showers measured with extended surface arrays equipped with scintillation counters and/or water tanks can be found in \cite{Mono,Artamonov,InclinedIzv,Zas,KCDC} and elsewhere.

In the rest of the paper, we focus on the Yakutsk array data. While the previous studies of zenith angle distribution, $f(\theta,E)$, were performed mainly at the highest energies where the full trigger efficiency is reached, our aim here is to elucidate the distribution in the wide energy range, with energy-dependent array exposure due to absorption of showers in the atmosphere. This makes it possible to use more data at lower energies in the analysis of CR arrival directions that were not involved previously.

\begin{figure*}[t]\centering
\includegraphics[width=\columnwidth]{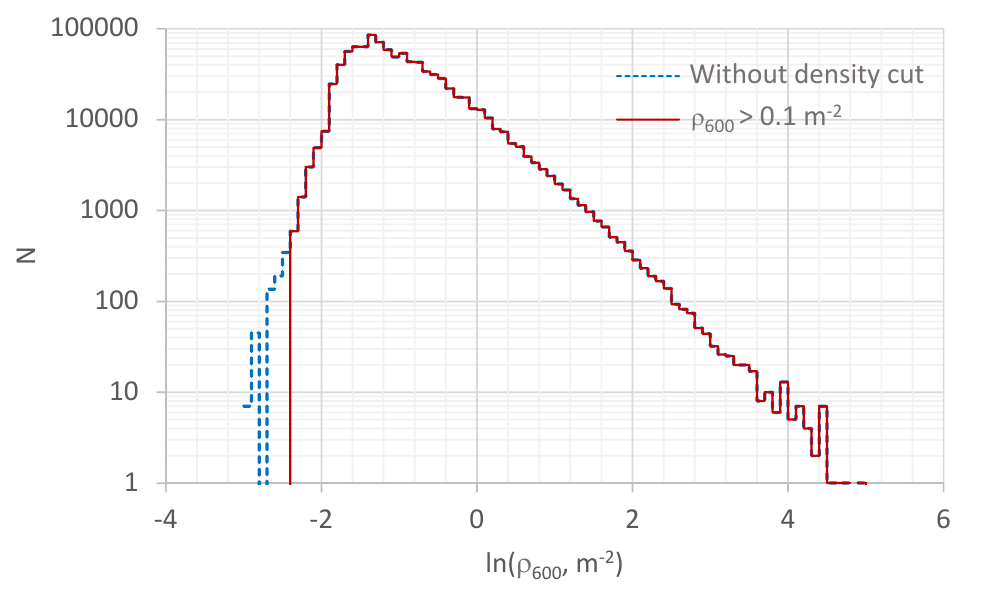}
\includegraphics[width=\columnwidth]{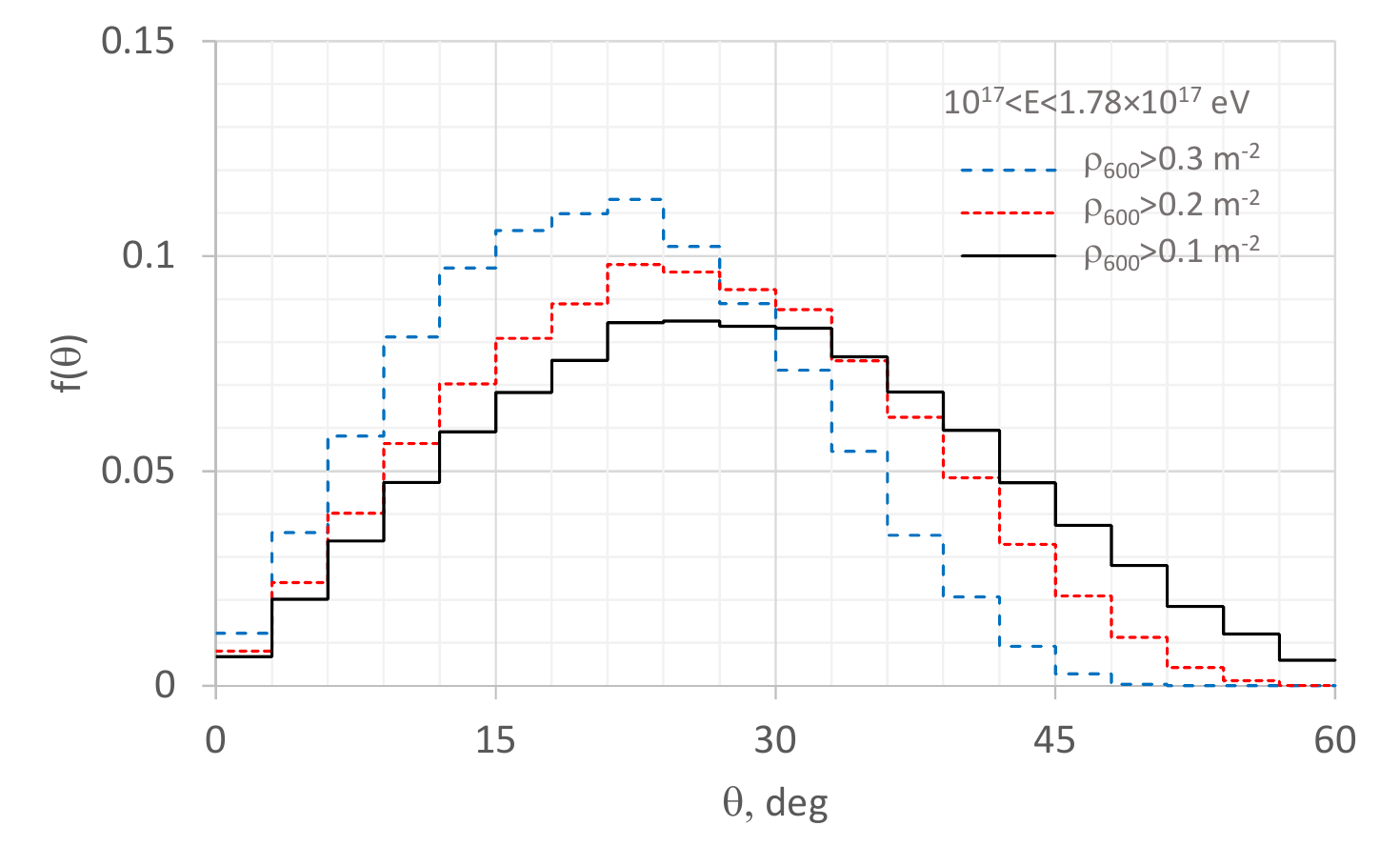}
  \caption{Left panel: Distribution of the measured $S_{600}$ parameter.
  Right panel: Zenith angle distribution of EAS event rate with different $S_{600}$ thresholds.}
\label{Fig:S600Distr}\end{figure*}

To simplify the treatment of the shower attenuation effects, in the following we use the threshold of particle density at 600 m from the shower core, $S_{600}$, equal to 0.1 m$^{-2}$, which is chosen to be well above the intrinsic instrumental thresholds of the array. A benefit of using this technique is \emph{a posteriori} selection of showers almost independently of shower core position within the array area. The $S_{600}$ distribution of showers detected with the Yakutsk array is given in the left panel of Fig.\ \ref{Fig:S600Distr} with and without the density threshold at 600 m, while the distributions $f(\theta,E)$ modified by the density thresholds are shown in the right panel. The figure demonstrates the influence of the $S_{600}$ threshold on the measured zenith angle distribution.

The paper is structured as follows. A brief description of the experimental equipment and the procedure of data acquisition and selection for analysis are given in Section II. Absorption of showers in the atmosphere is treated in Section III. The resultant zenith angle distribution is analyzed in Section IV. Finally, in Section V, the observed and expected distributions are applied in equatorial coordinates in order to test the null and alternative hypotheses of the distribution of arrival directions of CRs.

\section{The Yakutsk array. Data acquisition and selection for analysis}
The geographical coordinates of the Yakutsk array site (about $100$ m above sea level, or $x_0=1020$ g/cm$^2$) are $61.7^0$N, $129.4^0$E. The array is formed by 58 ground-based and four underground scintillation counters of charged particles (electrons and muons) supplemented by 48 detectors of atmospheric Cherenkov light consisting of photomultiplier tubes \cite{Mono,Kashiwa,NIM}.

Stations on the ground with approximately $500$ m of separation contain a pair of scintillation counters (2 m$^2$ each). The total area covered by the stations was $S\approx17$ km$^2$ in the period 1974--1990 and $S\approx10$ km$^2$ between 1990 and 2000 (stage II); currently it is 8.2 km$^2$ \cite{JETP2007,AnoVa}. The energy range of EAS investigations is $10^{15}$ to $10^{20}$ eV \cite{Pylos2004,MSU,FlorenceMass,SpectraConverged}.

EAS events are selected from the background using a two-level trigger for detector signals (particle density $\rho>0.5$ m$^2$). The first level involves the coincidence of signals from two scintillation counters at a station within 2 $\mu$s; the second level involves the coincidence of signals from at least three nearby stations within 40 $\mu$s \cite{Mono,JETP2007}.

Several algorithms have been developed in the Yakutsk array group to evaluate the energy of the primary particle initiating EAS \cite{CERN2003,EMcomponent,SpectrumCher,Masses}. In this paper, the most recent method proposed in \cite{Glushkov2017} is used. Namely, the primary energy is estimated as in the SIBYLL-2.1 model \cite{SIBYLL}:
\begin{equation}
E=(0.37\pm0.01)\times S_{600}(0)^{1.02},\text{EeV},
\label{Eq:PrEnBusan}
\end{equation}
where $S_{600}(0)$ is the particle density at 600 m from the shower core measured in a vertical shower. In inclined showers $S_{600}(\theta)$ is estimated using the constant intensity cuts method described in the next section.

The selected sample of the Yakutsk array data consists of EAS events detected in the period of January 1974 to June 2008 \cite{AnoVa} with axes within the stage II array area at energies above $0.1$ EeV ($=10^{17}$ eV) and zenith angles $\theta\in(0^0,60^0)$.

\section{Treatment of attenuation of shower parameters in the atmosphere: constant intensity cuts}
The attenuation of shower size and particle density, $\rho(r,\theta)$, at zenith angle $\theta$ and core distance $r$ has been studied in a number of ground-based experiments, most notably Haverah Park \cite{HP}, AGASA \cite{AGASA1995}, and KASCADE \cite{Kascade}. The \textit{constant intensity cuts} method is used in the papers cited and elsewhere to evaluate the attenuation length of EAS parameters in the atmosphere. A detailed description and possible applications of the method are given, for example, in \cite{Gaisser}.

In order to apply the method in analysis of the Yakutsk array data we have calculated the effective array area, $S_{eff}$, as a function of $S_{600}$ and zenith angle, and defined a fixed observational time duration and acceptance angle, $T\Omega$. The Monte Carlo method is implemented using fake showers with random axes within the array area, and the particle density is approximated by the lateral distribution function:
$$
\rho=S_{600}(\frac{r}{600})^{-a}(\frac{r_m+r}{r_m+600})^{a-b}
(\frac{2000+r}{2600})^{-g}, m^{-2},
$$
where $r_m$ is the Moliere radius; $b=b_1+b_2\cos\theta+b_3\lg S_{600}(E,\theta)$; constants $a, b_i$,  and $g$ are fitted with the experimental data \cite{Glushkov2000,AgePrm}. Normal fluctuations of densities are defined by the variance \cite{Mono}:
$$
D_\rho^{det}=\rho^2(\beta^2+\frac{1+\alpha^2}{\rho S_{det}\cos\theta}),
$$
where $\alpha=0.45$; $\beta=0.16$; $S_{det}$ is the detector area.

To select showers detected with the array, the EAS event selection steps described in Section II were simulated numerically. The actual configuration of the stage II array stations was used to trigger events by two-level coincidences of detector signals.

\begin{figure}[b]\centering
\includegraphics[width=\columnwidth]{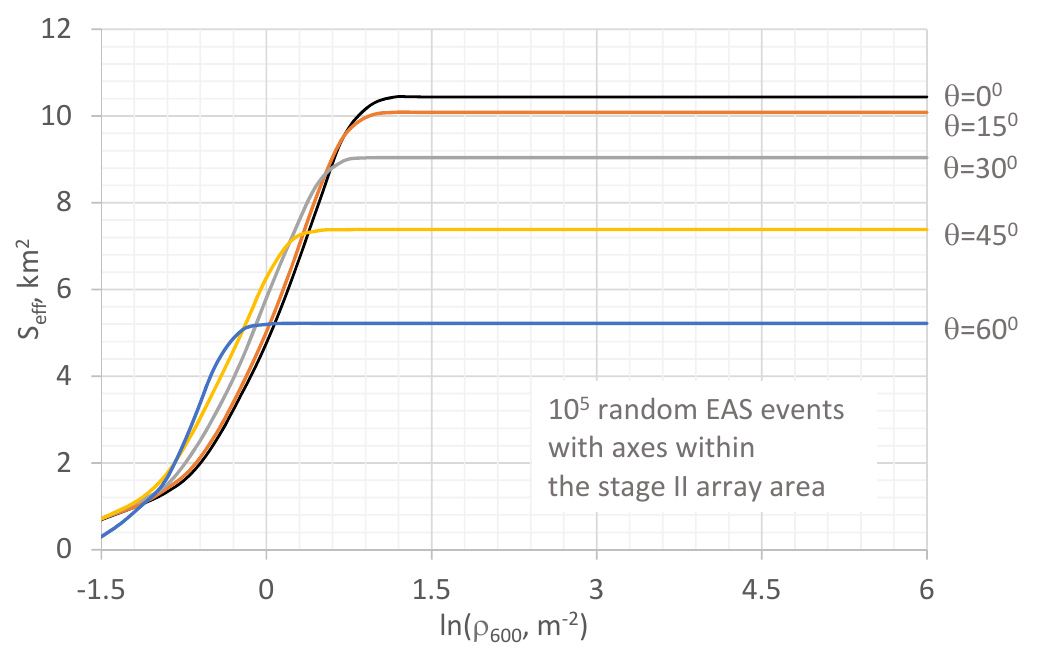}
  \caption{Effective area of the Yakutsk array vs $S_{600}$.}
\label{Fig:Seff}\end{figure}

The results are shown in Fig. \ref{Fig:Seff}. For showers with $S_{600}(E,\theta)>2.7$ m$^{-2}$, the effective array area is constant due to the 100\% trigger efficiency. For lower densities, attenuation in the atmosphere results in reduced trigger efficiency.

\begin{figure}[t]\centering
\includegraphics[width=0.94\columnwidth]{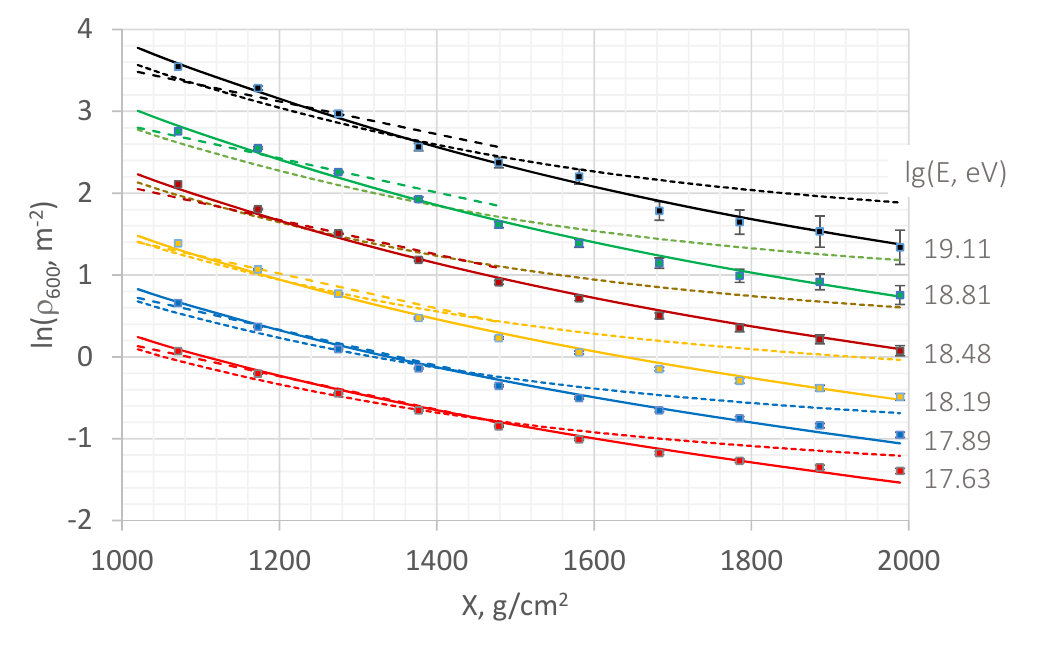}
  \caption{Constant intensity cuts: Attenuation of $S_{600}$ in the atmosphere. Experimental data are given by points parameterized with $\lg(E)$; functional approximations are shown by dots (\cite{Pravdin2007}), dashed lines (\cite{Glushkov2017}), and solid curves (this work).}
\label{Fig:CIC}\end{figure}

Constant intensity cuts of the observed $S_{600}$ spectra are then defined in zenith angle intervals at intensities $I=n(S_{600},\theta)/S_{eff}=const$, where $n(S_{600},\theta)$ is the number of showers. The resultant function $S_{600}(\theta)/_{I=const}$ is the attenuation curve of the density as a function of zenith angle, habitually in the form of $x=x_0\sec\theta$. The Yakutsk array data result in the $S_{600}(x)$ curves shown in Fig. \ref{Fig:CIC}, where constant intensities are substituted for primary energies connected to $S_{600}(x_0)$ by Eq. \ref{Eq:PrEnBusan}.

To fit the attenuation curves, the functional approximation is chosen which is formed by the weighted sum of two exponentials
\begin{equation}
S_{600}(x)=S_{600}(x_0)((1-\beta)e^{\frac{x_0-x}{\lambda_1}}+\beta e^{\frac{x_0-x}{\lambda_2}}),
\label{Eq:Atten}
\end{equation}
where $\lambda_1=230$ g/cm$^2$; $\lambda_2=900$ g/cm$^2$; $\beta=0.5S_{600}^{-0.2}(x_0)$. Previous approximations are shown as well.

As appears from the algorithm given above, the shape of the attenuation curve $S_{600}(x)$ is independent of the primary energy estimation; actually, it is parameterized by the measured $S_{600}(x_0)$.

\section{Zenith angle distribution of EAS event rate in the primary energy intervals}
This distribution is a result of shower absorption in the atmosphere as well as the arrival directions of the primaries. In turn, the absorption rate is linked to the threshold energy of showers, which depends on the particle density threshold of the detectors, shower core coordinates, and so on. As was stated in Section I, we replace all the factors by the threshold $S_{600}^{thr}=0.1$ m$^{-2}$, which leads to a predictable shower absorption effect.

\subsection{Measured zenith angle distributions and those expected for isotropy}
It was shown long ago that fluctuations of some shower parameters in a narrow energy bin, for example, shower sizes, $N_e$, $N_{\mu}$, and particle density, can be approximated at sea level by a log-normal distribution \cite{Dedenko,Kalmykov}. In particular, it was demonstrated with the experimental data of AGASA \cite{AGASA2003} and with a CORSIKA simulation of the scintillation counter signal \cite{Rubtsov} that $y=\ln(S_{600})$ in EAS events can be approximated by a Gaussian. Assuming an isotropic flux of CRs in the energy range of $(0.1,10)$ EeV, one can derive an analytic expression to describe the zenith angle distribution of showers that have survived after cutting at the particle density threshold and have reached detectors at sea level.

Constraints on the anisotropy of arrival directions in the range were set using harmonic analysis \cite{PravdinJETP,PravdinIzv,KASCADEHarm} and the South-North method \cite{IJMPDisotropy}. The stringent upper limit for the first harmonic amplitude was 0.1\% at a primary energy of $7\times10^{14}$ eV and 1.25\% at around $E\approx1$ EeV. This bounds our isotropic flux assumption. On the other hand, if there is actually anisotropy in CR arrival directions, it should result in a discrepancy between observed and expected zenith angle distributions; the issue will be discussed in the next section.

At the highest energies where $y>1$, the absorption of showers is negligible, $S_{eff}$ is constant, and the distribution is compatible with the `unabsorbed' $f(\theta)=\frac{4}{3}\sin(2\theta)$ \cite{Arrival,MinWidth}. It is formed by the dependence of the array acceptance on solid angle $d\Omega\propto\sin\theta$ and on zenith angle $\propto\cos\theta$.

\begin{figure}[t]\centering
\includegraphics[width=\columnwidth]{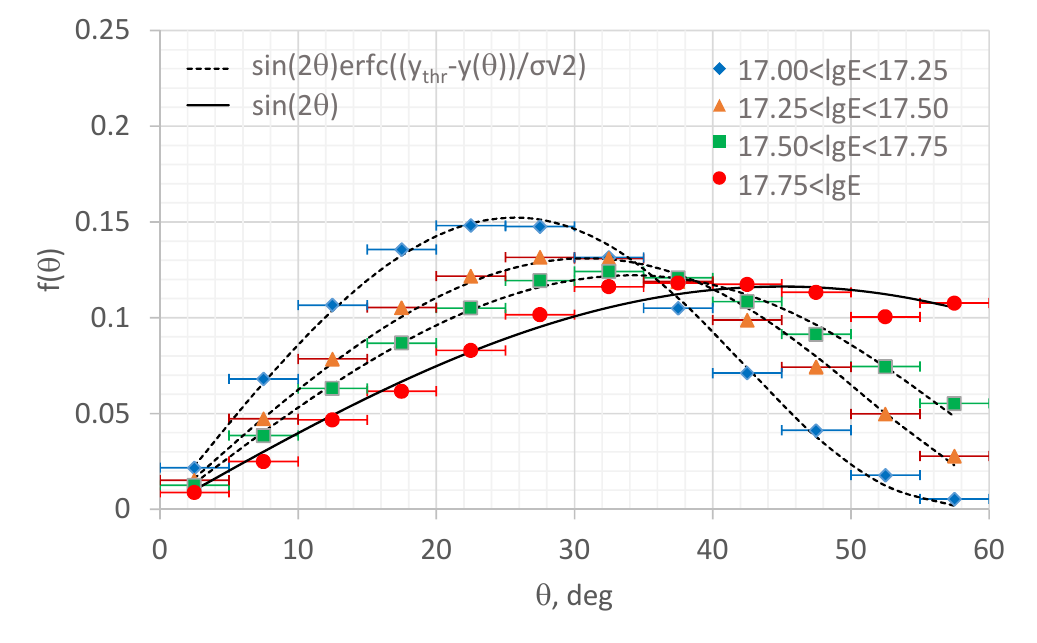}
  \caption{Zenith angle distribution of the EAS event rate observed with the Yakutsk array in energy intervals. Expected distributions are shown by curves.}
\label{Fig:Fit}\end{figure}

At lower energies, the particle density threshold cuts the right-hand tail of the distribution.
To form the clipped distribution expected for the Yakutsk array with log-normal fluctuations of $S_{600}$, one should add the cutting factor, which is the normal distribution stump:
\begin{equation}
f(\theta)=C\sin(2\theta)\textup{erfc}(\frac{y_{thr}-\overline{y(\theta)}}{\sqrt{2}\sigma}),
\label{Eq:clip}
\end{equation}
where $C$ is a normalizing constant; $y_{thr}=-2.303$; and $\sigma$ is the r.m.s. deviation.  The mean density is given by the attenuation curve (Fig. \ref{Fig:CIC}, Eq. \ref{Eq:Atten}). A parameter $\sigma$ in the formula is adjustable in order to fit the experimental data.

In Fig.\ \ref{Fig:Fit}, experimental zenith angle distributions ($S_{600}>0.1$ m$^{-2}$) are shown in four energy intervals together with the expected isotropic distribution~(\ref{Eq:clip}), where data histograms are represented by points in the middle of the bins for convenience.

\begin{figure}[t]\centering
\includegraphics[width=\columnwidth]{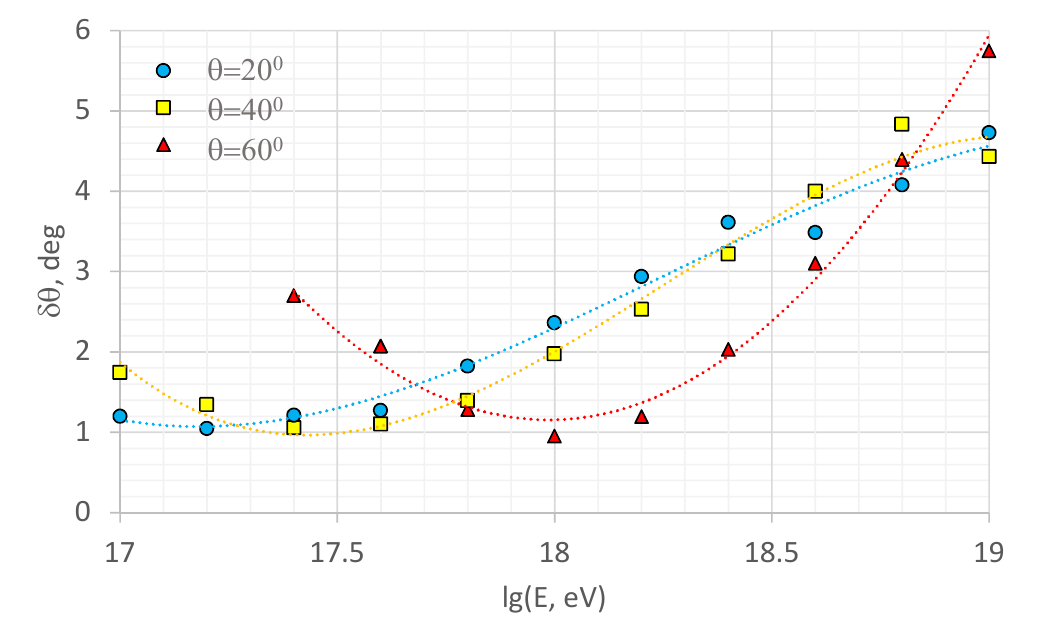}
  \caption{Uncertainty of zenith angle measurement in the Yakutsk array experiment. Calculated values are shown by points and smoothed by curves.}
\label{Fig:Accuracy}\end{figure}

\subsection{Influence of energy and angular uncertainty on zenith angle distribution measured in the Yakutsk array experiment}
The shower energy uncertainty was estimated for the Yakutsk array surface detectors by Pravdin et al.~\cite{Pravdin2007} in the energy interval $(10^{18},10^{19.5})$ eV, with the zenith angle $\theta\in(0^0,60^0)$, as $\delta E/E\in(0.15,0.35)$ (random errors), and $\delta E/E\in(0.25,0.3)$ (systematic errors). The maximum of the total relative error is 0.44.

A shift of energy leads to the distortion of the zenith angle distribution (illustrated in Fig.~\ref{Fig:Fit}), which can be roughly described by the maximum position, $\theta_{max}(E)$, shifting with energy. Measured distributions are used to estimate $d\theta_{max}/d\lg E=10^0\pm5^0$. The resultant uncertainty is found to be $\delta\theta_{max}=1.6^0\pm0.8^0$, well below the typical angular error of the Yakutsk array~\cite{Pravdin2007}.

The EAS arrival direction is evaluated in the Yakutsk array group using time delays of the first scintillation signals from the shower in stations. Zenith and azimuth angles are found by the least squares optimization in the plane approximation of the shower front. The accuracy of time delays is determined by the synchronization system of detectors, providing 100 ns time slicing. The algorithm is insensitive to neither energy nor the mass composition of primaries.

We have estimated the zenith angle uncertainty utilizing the same program as was used to calculate $S_{eff}$ of the array. The shower front curvature measured by Knurenko et al.~\cite{Curvature} at energies above $10^{16}$ eV was used to calculate the time delays between stations. The results are given in Fig.~\ref{Fig:Accuracy}. The dependence of the accuracy on zenith angle is due to $L\cos\theta$, where $L$ is the distance between stations, while the dependence on energy is determined by the distant stations (more delayed signals because of front curvature) included in shower detection at higher energies.

Our estimation of the zenith angle uncertainty is in qualitative agreement with previous results, for example \cite{Pravdin2007}, but demonstrates different energy dependence owing to the contribution of the shower front curvature to the angular uncertainty. Only a large scale structure of zenith angle distribution of EAS event rate can be analyzed using the Yakutsk array data at angular sizes greater than $6^0$.

The energy bin width is limited by the CR mass composition changing with energy. It can be estimated from Fig.~3 in Sabourov et al.~\cite{Sabourov}: $d\lg(A)/d\lg(E)=0.8\pm0.2$ within the interval $(0.1,10)$ EeV. According to superposition approximation, EAS parameters depend on the primary energy and mass in combination $f(E/A)$. In other words, the relative variation in mass composition is equivalent to the variation of the primary energy. As a consequence, the energy bin width, $\Delta\lg(E)$, should be less than 0.75 in order to be in accordance with the zenith angle uncertainty $\delta\theta<6^0$.

\begin{figure}[t]\centering
\includegraphics[width=\columnwidth]{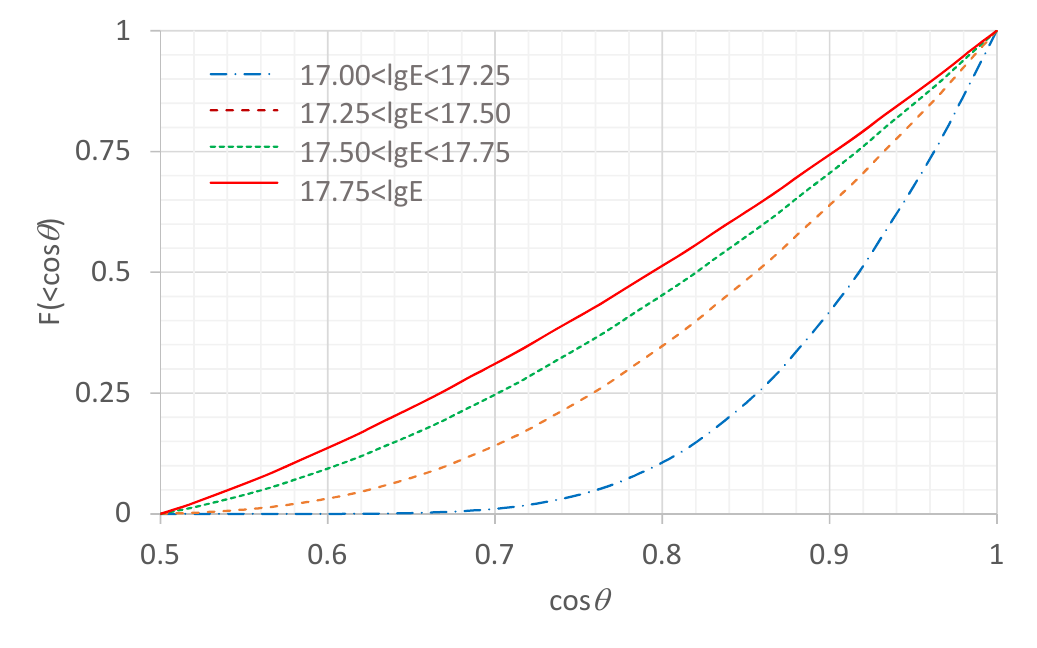}
  \caption{Numerical approximation of integral distributions of zenith angle in energy intervals.}
\label{Fig:Int}\end{figure}

\begin{figure*}[t]\centering
\includegraphics[width=1.15\columnwidth]{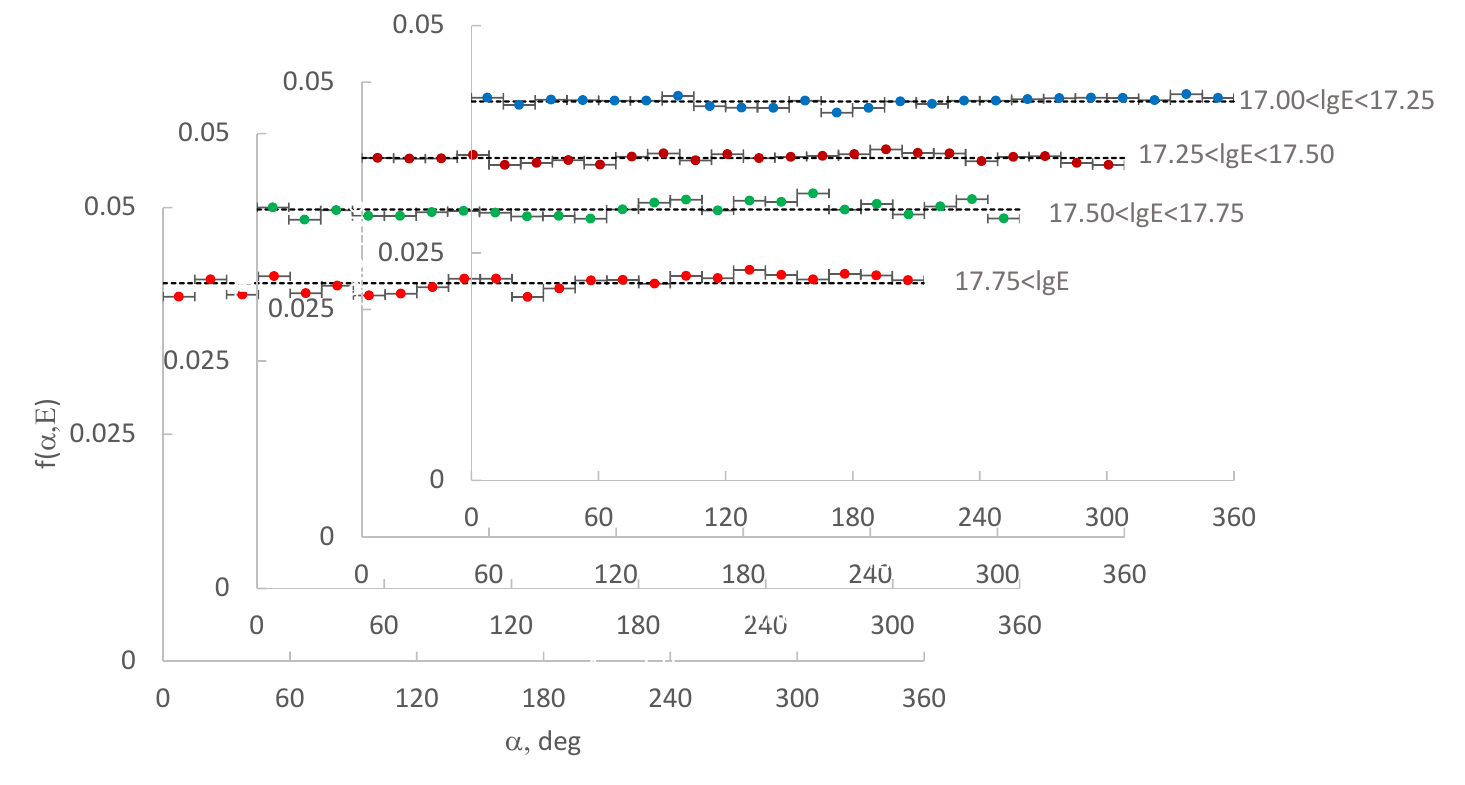}
\includegraphics[width=0.9\columnwidth]{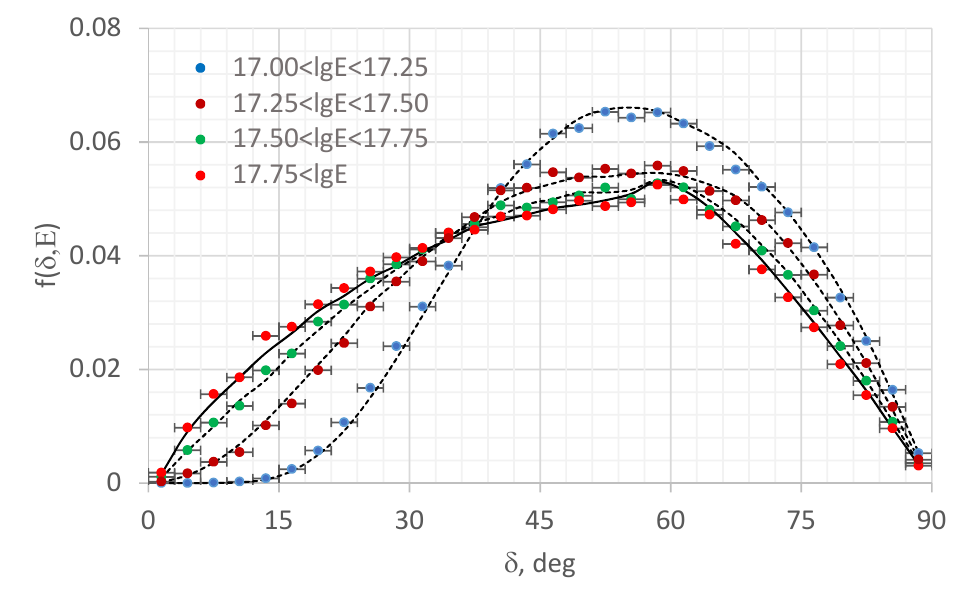}
  \caption{Left panel: Right ascension distribution of CR arrival directions.
  Right panel: Declination distribution of CR arrival directions.
  The Yakutsk array data in energy intervals are shown by points (horizontal bars indicate zenith angle bins); expected-for-isotropy distributions are shown by curves derived from the uniform and zenith angle (\ref{Eq:clip}) distributions.}
\label{Fig:Qtrl}\end{figure*}

\section{Application to arrival directions of cosmic rays in equatorial coordinates}
The zenith angle distribution (\ref{Eq:clip}) together with a uniform azimuth distribution can be used to check whether the observed distribution of CR arrival directions in a given energy interval is isotropic or not. In order to do so, it is convenient to transform directions from horizontal to equatorial coordinates.

\subsection{Comparing measured and expected arrival directions}
A straightforward way to produce isotropic arrival directions as formed by the surface detector angular acceptance function in equatorial coordinates - right ascension, $\alpha$, and declination, $\delta$, is to generate a random zenith angle [$\theta$, from the distribution (\ref{Eq:clip})], azimuth ($\phi$, from a uniform distribution), and sidereal time ($st$, from a uniform distribution) and then transform it to $(\alpha,\delta)$. Previously, the method used to be applied \cite{Mono,MinWidth} with a zenith angle distribution of unabsorbed showers.

Here, a random azimuth angle and sidereal time are generated in the same way. To generate a random zenith angle, however, a well-known inverse cumulative distribution function technique~\cite{MonteCarlo} is implemented, namely, the function $F(<\theta)$ numerically derived from (\ref{Eq:clip}) is used (Fig.~\ref{Fig:Int}). Uniform pseudorandom numbers $R_i=F$ generated by a multiplicative method in the interval $(0,1)$ are then transformed to $\cos(\theta_i)$, interpolating the inverse function.

\begin{table*}[b]
\begin{center}
\caption{Application of the $\chi^2$-statistical goodness of fit test to distributions in equatorial coordinates
observed with the Yakutsk array and those expected-for-isotropy. A critical point of the probability is assumed to be 0.01.}
\begin{tabular*}{0.7\textwidth}{@{\extracolsep{\fill}}lrrrrrrrr}
\hline\hline
Energy bins, &    $N$  &\multicolumn{3}{c}{$\delta$ distribution} &\multicolumn{3}{c}{$\alpha$ distribution} \\
\cline{3-5}\cline{6-8}
$\lg(E, eV)$ &         & $\chi^2$  & $\nu$ & $P(\geq\chi^2)$,\% & $\chi^2$  & $\nu$ & $P(\geq\chi^2)$,\%\\\hline
 17.00-17.25 & 326775  &    362.01 &  5    & $<1.00$    &     11.63 &  7    & 11.35         \\
 17.25-17.50 & 207736  &    110.74 &  5    & $<1.00$    &     32.99 &  7    & $<1.00$       \\
 17.50-17.75 & 106438  &     65.79 &  5    & $<1.00$    &     21.93 &  7    & $<1.00$       \\
 17.75-18.00 &  46527  &     17.56 &  5    & $<1.00$    &      8.25 &  7    & 31.07         \\
 18.00-18.25 &  17837  &     10.25 &  5    &  6.85      &      7.11 &  7    & 41.78         \\
 18.25-18.50 &   6369  &     13.91 &  5    &  1.62      &      6.66 &  7    & 46.56         \\
 18.50-18.75 &   2178  &      9.32 &  5    &  9.69      &      7.15 &  7    & 41.38         \\
 18.75-19.00 &   1005  &      8.89 &  5    & 11.35      &     14.39 &  7    &  4.46         \\
\hline\hline
\end{tabular*}
\end{center}
\end{table*}

The resultant `isotropic' distribution of arrival directions in energy intervals is shown in Fig.~\ref{Fig:Qtrl} in comparison with observational data of the Yakutsk array in equatorial coordinates. The uniform $f(\alpha)$ distribution is smoothed down by the Earth's rotation, while the declination distribution is a result of the array exposure applied to the event rate of detected showers. Distributions at energies above $10^{17.75}$ eV are put together to emphasize the same expected distribution of showers in declination.

The degrees of consistency between observed and isotropic equatorial distributions in energy intervals are illustrated in Table I. Pearson's chi-square goodness of fit test is used to calculate the probability value, $P(\geq\chi^2)$, that the statistic with $\nu$ degrees of freedom would be the same as or greater than the value inherent in random fluctuations.

In the energy region above $10^{18}$ eV, the observed distribution of arrival directions is in agreement with isotropic expectation, in harmony with previous results \cite{AnoVa,IJMPDisotropy}, while in the interval $(10^{17},10^{18})$ eV there is an explicit deviation from isotropy. Pravdin et al. \cite{PravdinJETP} revealed that the effect is a result of irregular observation conditions and seasonal variations of the EAS event rate. More detailed analysis is required to distinguish a real anisotropy of astrophysical origin from seasonal and observational artifacts.

\subsection{Harmonic analysis in 2D equatorial map}
The uniform right ascension distribution dictates the application of the harmonic analysis, where all harmonic amplitudes are zero in the case of isotropic distribution of arrival directions, assumed as a null hypothesis, $H_0$, and the first harmonic amplitude becomes non-zero if there is a source of CRs~\cite{Kras}. The phase of the first harmonic points to the source position.

The main disadvantage of the method is its constraint to one-dimensional uni- or bimodal distribution in right ascension. A series expansion of the declination distribution consists of the set of non-zero amplitudes all of which must be taken into account. Consequently, spherical harmonics cannot be straightforwardly applied to equatorial coordinates in a 2D map of arrival directions.

An appropriate approach is to use one-dimensional analysis in a sliced map, that is, the right ascension rings in declination bins, for example $\alpha\in(0,360^0)$, $15^0i\leq\delta<15^0(i+1)$, where $i=0,..,5$. In this case, the declination distribution determines the relative number of CRs falling into rings and the resolving power of the array in a declination bin. Particularly, the Yakutsk array is blind to declinations below $\delta=1.7^0$ if showers are detected at $\theta<60^0$.

\begin{table*}[b]
\begin{center}
\caption{Ratio of the first harmonic amplitudes, observed and isotropic, $r=A_1/A_1^{iso}$, and the chance probability, $P(>A_1)$\%, derived from the Yakutsk array data in $\alpha$-rings.}
\begin{tabular*}{1.0\textwidth}{@{\extracolsep{\fill}}lrrrrrrrrrrrrrr}
\hline\hline
Energy bins,&\multicolumn{2}{c}{$\delta\in(0^0,90^0)$}&\multicolumn{2}{c}{$\delta\in(0^0,15^0)$}
&\multicolumn{2}{c}{$\delta\in(15^0,30^0)$}&\multicolumn{2}{c}{$\delta\in(30^0,45^0)$}
&\multicolumn{2}{c}{$\delta\in(45^0,60^0)$}&\multicolumn{2}{c}{$\delta\in(60^0,75^0)$}
&\multicolumn{2}{c}{$\delta\in(75^0,90^0)$}\\
\cline{2-3}\cline{4-5}\cline{6-7}\cline{8-9}\cline{10-11}\cline{12-13}\cline{14-15}
$\lg(E,eV)$&$r$&$P(>A_1)$&$r$&$P(>A_1)$&$r$&$P(>A_1)$&$r$&$P(>A_1)$&$r$&$P(>A_1)$&$r$&$P(>A_1)$&$r$&$P(>A_1)$\\\hline
  17.0-17.5& 4.45& 0.17E-04&1.56& 14.65 & 1.28& 27.67 & 2.35&  1.30 & 5.11& 0.13E-06  & 1.97&  4.75& 0.39&88.50\\
  17.5-18.0& 3.56& 0.46E-02&1.15& 35.24 & 1.37& 22.74 & 2.52&  0.68 & 3.66& 0.28E-02  & 0.49& 82.65& 0.92&51.47\\
  18.0-18.5& 1.55&   15.34 &1.75&  9.06 & 0.95& 49.42 & 0.82& 58.65 & 0.92&    51.41  & 1.14& 36.19& 1.35&23.83\\
  18.5-19.0& 1.36&   23.32 &1.28& 27.46 & 0.54& 79.87 & 0.55& 78.96 & 2.05&     3.71  & 1.55& 15.32& 0.79&61.23\\
\hline\hline
\end{tabular*}
\end{center}
\end{table*}

\subsubsection{Testing the null hypothesis}
 The formalism of harmonic analysis in $\alpha$-rings is the same as in previous studies (e.g. \cite{AnoVa}, and references therein); namely, the amplitude and phase of the $k$-th harmonic are calculated by formulating the right ascension distribution as a sum of delta functions $\sum_{i=1}^N\delta(\alpha-\alpha_i)$:

$A_k=\sqrt{a_k^2+b_k^2}$; $\phi_k=\arctan(b_k/a_k)$,

where $a_k=\frac{2}{N}\Sigma_i\cos(k\alpha_i)$; $b_k=\frac{2}{N}\Sigma_i\sin(k\alpha_i)$. The only difference is in the sorting of arrival directions into declination bins.

\begin{figure}[t]\centering
\includegraphics[width=\columnwidth]{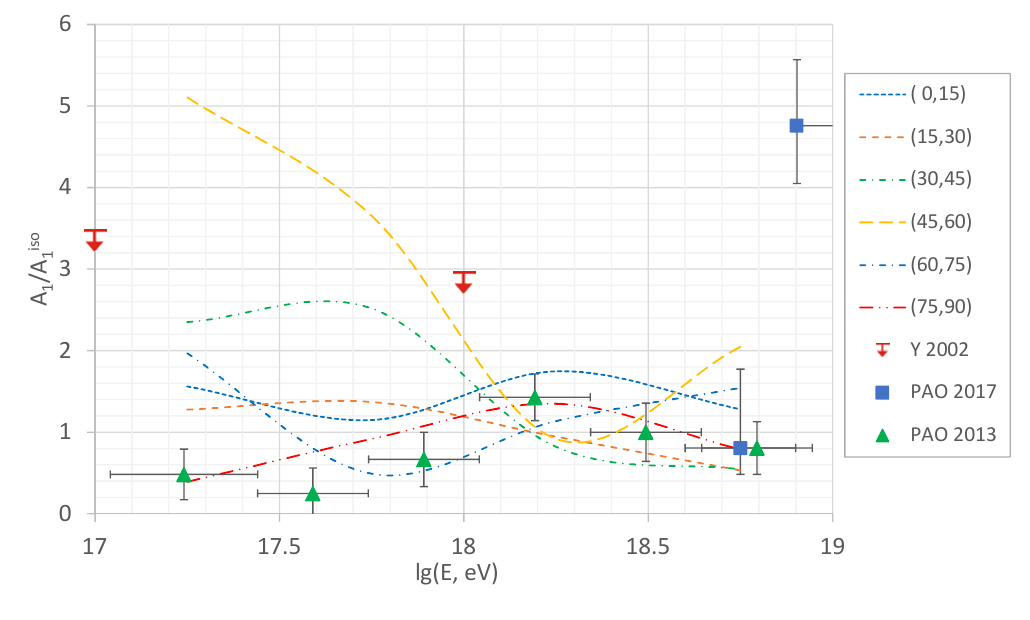}
  \caption{Comparison of the first harmonic amplitudes of the right ascension distribution of CRs measured in the Pierre Auger Observatory and the Yakutsk array experiments. Ratio of observed to expected-for-isotropy amplitudes are derived in declination intervals: $-90^0<\delta<25^0$ (PAO 2013 \cite{PAO2013}); $-90^0<\delta<45^0$ (PAO 2017 \cite{PAO2017}); $0^0<\delta<90^0$ (Yakutsk 2002 \cite{PravdinIzv}); results of this work (from Table II, smoothed by curves) are denoted by the declination intervals, in degrees, to where they belong.}
\label{Fig:Amplitudes}\end{figure}

The resultant first harmonic amplitude in units of the isotropic amplitude, $A_1^{iso}=\sqrt{\pi/N}$ \cite{AnoVa}, and the probability that an isotropic distribution will produce $A_1$ greater than or equal to the observed amplitude by chance is given in Table II in energy and declination bins.

The null hypothesis cannot be rejected in any bin above $10^{18}$ eV because the chance probability is greater than the critical value of 1\%. However, in a declination bin $(45^0,60^0)$ at energies within $(10^{17},10^{18})$ eV, there is a significant deviation from isotropy in qualitative agreement with the $\chi^2$ test results in Table I. As a consequence, conventional harmonic analysis in right ascension with integrated declinations [marked `$\delta\in(0^0,90^0)$' in Table II] exhibits a deviation from isotropy, too.

To distinguish the instrumental and seasonal origin of the effect from astrophysical sources, one can divide the dataset into seasonal subsets and test the phase stability  \cite{PravdinJETP}. The result is presented in Table III for a declination bin where the deviation from isotropy is located. Seasonal variation of the phase is evident, so the observed anisotropy cannot be attributed to an extraterrestrial source with fixed angular coordinates.

\begin{table}[b]
\begin{center}
\caption{Seasonal dependence of the first harmonic phase (right ascension, degree).
Declinations are within the interval $(45^0,60^0)$.}
\begin{tabular*}{0.48\textwidth}{@{\extracolsep{\fill}}lrrrr}
\hline\hline
$\lg(E, eV)$&17.0-17.25&17.25-17.5&17.5-17.75&17.75-18.0\\\hline
    Spring  & -113.9   &  -97.5   &  -122.3  & -119.6   \\
    Summer  &  -47.2   &  -86.6   &   -54.1  &    5.5   \\
    Autumn  &   31.2   &  -61.7   &     5.8  &    3.8   \\
    Winter  &  168.0   & -145.7   &  -166.1  & -155.1   \\
\hline\hline
\end{tabular*}
\end{center}
\end{table}


Previous results of the harmonic analysis in right ascension were taken with CR arrival directions integrated over the whole observable declination range because of unknown zenith angle/declination distribution of absorbed showers. The first harmonic amplitudes available in the energy range $0.1$ to $10$ EeV are shown in Fig. \ref{Fig:Amplitudes}.

Pravdin et al. \cite{PravdinIzv} set an upper limit on $A_1$ considering an anti-sidereal time vector caused by seasonal variations of the EAS event rate. Our results are in agreement with these limits except for the declination interval $(45^0,60^0)$ discussed above.

In spite of the different observable sky regions, the data of PAO and the Yakutsk array demonstrate the first harmonic amplitudes which do not significantly exceed the isotropic expectation at energies below $8$ EeV \cite{PAO2013}. At higher energies, however, the PAO collaboration revealed a large-scale anisotropy in the arrival directions of CR: $A_1/A_1^{iso}=4.8$ at more than the 5.2$\sigma$ level of significance \cite{PAO2017}. Unfortunately, the anisotropy dipole points to the declination $\delta=-24^{+24}_{-13}$ degrees, invisible for the Yakutsk array.

\begin{figure}[t]\centering
\includegraphics[width=\columnwidth]{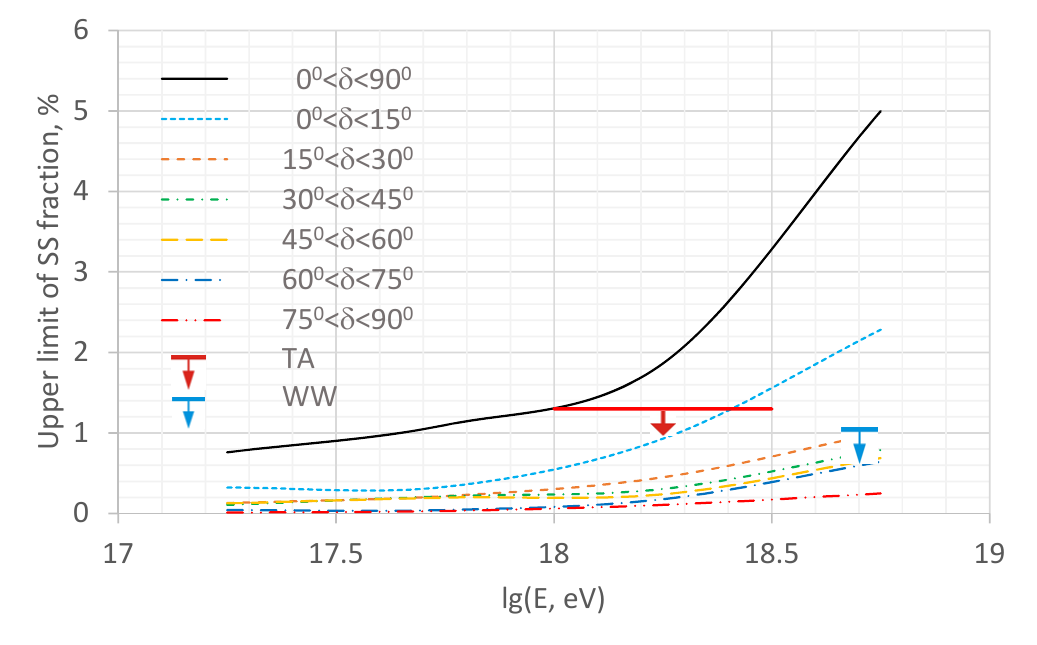}
  \caption{Upper limit of SS fraction in declination bins. Telescope Array's upper limit on the fraction of EeV protons of galactic origin (TA, \cite{TA}) and Wibig \& Wolfendale's upper limit on the fraction of galactic light nuclei (WW, \cite{WW}) in the CR beam are given for comparison.}
\label{Fig:Limit}\end{figure}

In general, there is no statistically significant deviation from isotropy in the arrival directions of CRs exceeding instrumental and seasonal effects in all energy and declination bins observed by the Yakutsk array, so the null hypothesis is not rejected by the data analyzed.


In this approach, however, it is not possible to draw a conclusion about the statistical power of the method to reject $H_0$ if a source of CRs has actually contributed to the dataset. An alternative hypothesis is needed to contrast with $H_0$.

\subsubsection{Alternative hypothesis: a separable source in the uniform background}
One possible unimodal distribution model from a variety of alternative hypotheses, $H_1$, is that consisting of a separable source of CRs, $SS$, giving a fraction $f_{SS}$, while all other sources form the uniform background providing a fraction $1-f_{SS}$ of the total flux.

The mass composition of astroparticles from the hypothesized $SS$ is constrained by deflections in the galactic magnetic field. Because of the critical energy \footnote{where the Larmor radius equals the coherence length of the turbulent component of the galactic magnetic field.} of charged particles $E_C\sim 0.3$ EeV \cite{CritEnergy}, it is thought to be neutral particles (photons? neutrinos?), if a source is not nearby. At energies above $1$ EeV, however, the magnetic dispersion of protons is less than the declination bin width $15^0$, so protons may be supposed to be emitted, too.

Comparing the first harmonic amplitude under $H_1$ with the measured $A_1^{exp}$ one can calculate a probability. Specifically, the probability $P(A_1\leq A_1^{exp})$ below the critical value means that $H_1$ can be rejected for a given fraction $f_{SS}$.

The first harmonic amplitude is calculated in $H_1$ as a sum of $f_{SS}N$ vectors of the length $2/N$ pointing to a source direction $\alpha_{SS}$, and $(1-f_{SS})N$ isotropic vectors of the same length. According to the central limit theorem, the isotropic summand is a circular Gaussian with $A_{iso}^2=4(1-f_{SS})/N$ \cite{MinWidth}. Averaging a sum $A_1^2=(2f_{SS}+A_{iso}\cos\alpha)^2+A_{iso}^2\sin^2\alpha$ over right ascension, one can get $\overline{A_1^2}=4f_{SS}^2+4(1-f_{SS})/N$.

If the alternative hypothesis is actually implemented in reality, and the amplitude $A_{H1}^2=4f_{SS}^2+4(1-f_{SS})/N$ is realized in the entire CR population, then two consequences are possible: a) the null hypothesis can be rejected, where $\exp(-NA_{H1}^2/4)<0.01$ (estimating the statistical power of the Rayleigh test \cite{IJMPDisotropy}); b) an upper limit on the SS fraction can be set where the observed amplitude $A_1^{exp}$ is significantly less than the expected amplitude $A_{H1}$.

In this paper, the second approach is employed. The probability $P(A_1\leq A_1^{exp})$ under the alternative hypothesis is calculated using the Monte Carlo method. Isotropic $(1-f_{SS})N$ points and $f_{SS}N$ in $\alpha_{SS}$, where $N$ is the number of CRs detected in a particular $\alpha$-ring, are sampled $M=10^5$ times to calculate $A_1$. An upper limit $f_{SS}^{thr}$ is determined, where the probability is equal to $P_{crit}=0.01$, so the SS fraction above the threshold can be rejected.

Because the exposure varies with declination, the array-resolving power of SS contribution to the CR flux differs between declination bins. Additionally, the fraction of the total number of CRs emitted by all sources pointing to the spherical array with cross section $S_{eff}$, detected in the $i$-th ring, is simulated taking into account the array exposure. The resultant ratio of the number of CRs detected in all rings to the number of emitted CRs is found to be 0.187 in a diurnal cycle. In a similar way, a CR fraction detected in the $i$-th ring is calculated for a source supposed to be situated in the $i$-th declination bin. This ratio rises from $0.1$ to $0.87$ with declination.

The resultant upper limit of the SS fraction as a function of energy is shown in Fig. \ref{Fig:Limit}. Here, the data points are set at the center of four $\Delta\lg E=0.5$ energy bins, and are smoothed out by curves. The energy dependence can be explained by the decreasing number of detected EAS events as the energy is higher. The declination dependence is due to the irregular array exposure of SS supposed to lurk in a particular $\delta$-bin.

As a reference, an upper limit on the fraction of EeV protons of galactic origin derived by the Telescope Array collaboration \cite{TA}, and an upper limit on the fraction of galactic light nuclei in the CR beam set by Wolfendale \& Wibig \cite{WW} are shown. These limits are comparable `in average' to the conventional harmonic analysis result (marked `$0^0<\delta<90^0$') and the SS fraction limit in the declination interval $0^0<\delta<15^0$. In the case of conventional analysis, a separable source may be anywhere in the range $\delta\in(0^0,90^0)$, so the SS fraction limit is larger for the source luminosity. Finally, if the source location turns out to be within $\delta\in(-90^0,1.7^0)$ then there is no upper limit at all.

\section{Conclusions}
The zenith angle distribution of the EAS event rate detected with the Yakutsk array is reanalyzed, implementing an $S_{600}$ threshold instead of a multitude of detector density thresholds. The aim is to demonstrate a connection between the shower absorption rate in the atmosphere and $f(\theta)$ by comparing the observed and expected distributions.

The expected distribution is derived by assuming isotropic arrival directions of CRs and log-normal fluctuations of $S_{600}$. It is shown that, indeed, modification of the shape of the zenith angle distribution with CR energy is caused by the shower absorption and fluctuations. A simple analytic expression is proposed to calculate the expected $f(\theta)$, where the mean $S_{600}(x)$ attenuation is evaluated by the constant intensity cuts method.

A detailed comparison of measured and expected distributions is undertaken in order to check whether there is a statistically significant deviation from isotropy in the CR arrival directions or not. For this, uniform distributions of the sidereal time and azimuthal angle before transforming from the horizontal to the equatorial system  are implemented. Harmonic analysis in $\alpha$-rings bordered within $\delta$ bins is applied on a 2D equatorial map. As a result, above $10^{18}$ eV, the null hypothesis cannot be rejected in conformity with previous analyses of the Yakutsk array data. Below this threshold, in the energy interval $(10^{17},10^{18})$ eV, there is a significant deviation from uniformity in the declination bin $\delta\in(45^0,60^0)$, but the effect is a consequence of instrumental and seasonal variations of the array exposure and cannot be connected with a CR source located somewhere in equatorial coordinates.

By making use of an alternative hypothesis containing a separable source in an otherwise isotropic set of CR sources, the stringent upper limit on the fraction of the total CR flux from such a source is set, at least, in declination intervals above $\delta=15^0$.

\begin{acknowledgments}
The author is grateful to the Yakutsk array group for data acquisition and analysis. This work is supported in part by SB RAS (program II.2�/II.16-1) and RFBR (grant no. 16-29-13019).
\end{acknowledgments}


\end{document}